\documentclass[aps,prl,twocolumn,amsfonts,amsmath,floatfix,showpacs,superscriptaddress,footnoteintext]{revtex4}

\usepackage[final]{graphicx}
\usepackage{color}

\newcommand{\be}{\begin{equation}} \newcommand{\ee}{\end{equation}}

\newcommand{\br}{\mathbf{r}}
\newcommand{\bF}{\mathbf{F}}
\newcommand{\bG}{\mathbf{G}}
\begin{document}

\title{Reversing the Critical Casimir force by shape deformation}

\date{\today}

\author{Giuseppe Bimonte}
\affiliation{Dipartimento di Scienze
Fisiche, Universit{\`a} di Napoli Federico II, Complesso Universitario
MSA, Via Cintia, I-80126 Napoli, Italy}
\affiliation{INFN Sezione di
Napoli, I-80126 Napoli, Italy }

\author{Thorsten Emig}
\affiliation{Laboratoire de Physique
Th\'eorique et Mod\`eles Statistiques, CNRS UMR 8626, B\^at.~100,
Universit\'e Paris-Sud, 91405 Orsay cedex, France}

\author{Mehran Kardar} \affiliation{Massachusetts Institute of
Technology, Department of Physics, Cambridge, Massachusetts 02139, USA}

\begin{abstract}
  The exact critical Casimir force between periodically deformed
  boundaries of a 2D semi-infinite strip is obtained for conformally
  invariant classical systems.  Only two parameters (conformal charge
  and scaling dimension of a boundary changing operator), along with
  the solution of an electrostatic problem, determine the Casimir
  force, rendering the theory practically applicable to any shape and
  arrangement.  The attraction between any two mirror symmetric
  objects follows directly from our general result.  The possibility
  of purely shape induced reversal of the force, as well as occurrence
  of stable equilibrium points, is demonstrated for certain
  conformally invariant models, including the tricritical Ising model.
\end{abstract}


\maketitle

Fluctuation-induced forces (FIF) are ubiquitous in nature~\cite{Kardar:1999a};
prominent examples include van der Waals~\cite{Parsegian:2005ly}, 
and closely related Casimir forces~\cite{Casimir:1948bh,Bordag:2009ve},
originating from quantum fluctuations of the electromagnetic field. 
Thermal fluctuations in soft matter also lead to FIF, most pronounced near a
critical point where correlation lengths are large~\cite{Gennes:1978qf,Krech:1994kl}. 
Controlling the sign of FIF (attractive or repulsive) is important to myriad
applications in design and manipulation of micron scale devices.
This has been achieved with judicious choice of materials in case of 
quantum electrodynamic (QED) Casimir forces~\cite{Munday:2009},
and with appropriate boundary conditions for critical FIF~\cite{Soyka:2008,Hertlein:2008}.

The non-additive character of FIF has also prompted a quest for
reversing the sign of Casimir forces solely by manipulation of shapes.
The original impetus comes from the intriguing result by Boyer~\cite{Boyer:1968}
for the modification of QED zero point energy by a spherical metal shell. 
The suggestion that this result may imply repulsion between two hemispheres
was later ruled out by a general theorem for attraction between
mirror symmetric shapes~\cite{Kenneth:2006wn,Bachas:2007vz}.
There are indeed specific geometrical arrangements in which the normally
attractive QED force in vacuum appears repulsive when constrained
along a specific axis (e.g.~\cite{Levin:2010}), but is unstable when moved off such axis.
Indeed, a generalized Earnshaw's theorem for FIF in QED rules out the possibility
of stable levitation (and consequently force reversals) in most cases~\cite{Rahi:2010yg}.

\begin{figure}[h]
\includegraphics[width= 1.\columnwidth]{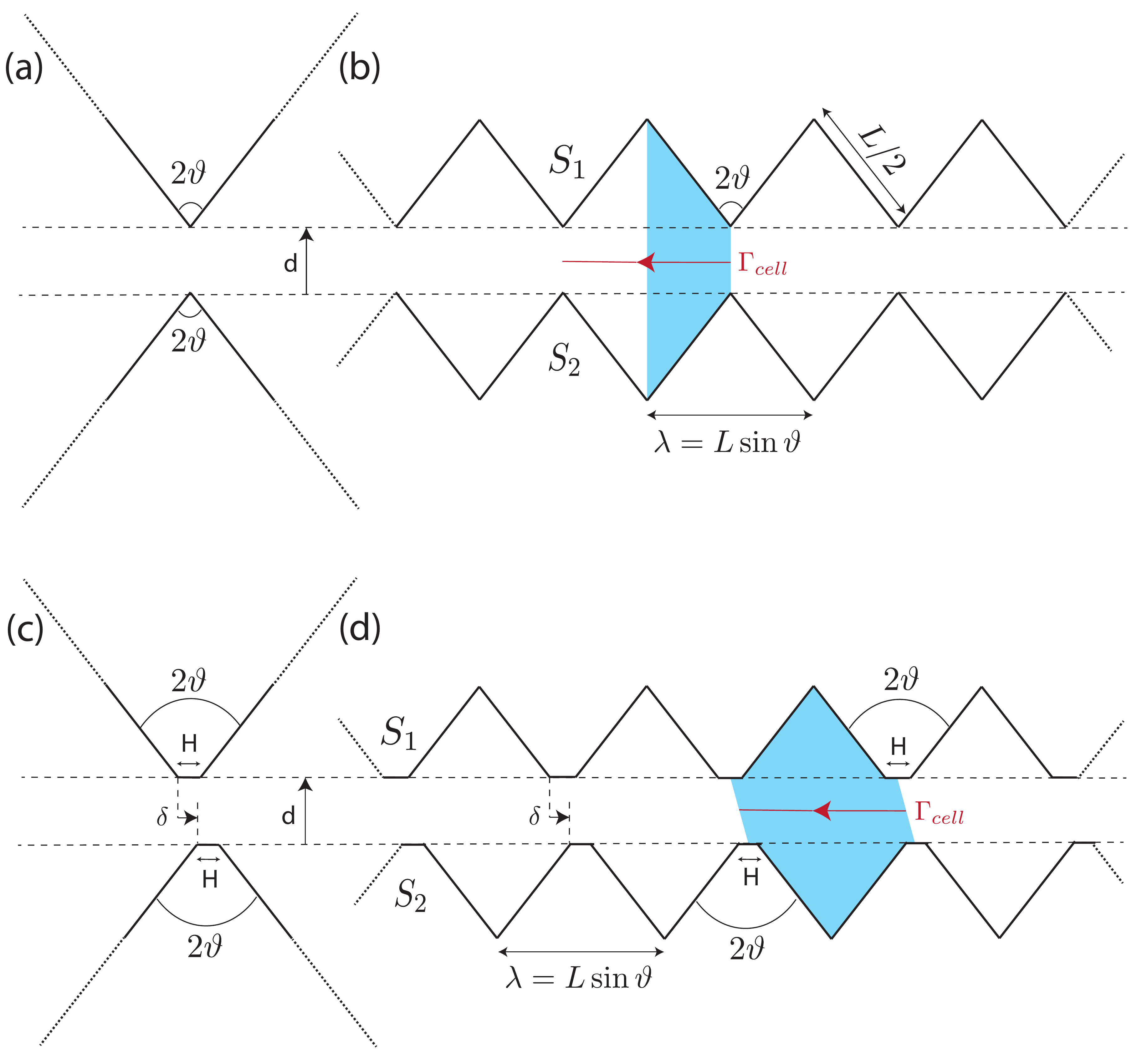}
\caption{Shapes considered: (a) two
  wedges, (b) strip with triangular corrugations, (c)  truncated
  wedges with lateral shift, (d) strip with truncated 
  corrugations and lateral shift. The blue regions mark half  a unit
  cell (b) and a full unit cell (d).}
\label{fig:geometry}
\end{figure}

Two dimensional (2D) membranes have provided yet another arena for
investigation of FIF, mostly focused on interactions arising due to
modifications of capillary fluctuations (see,
e.g.~\cite{Deserno:2012,Noruzifar:2013} and references therein).  More
recently, motivated by the possibility that the lipid mixtures
composing biological membranes are poised at
criticality~\cite{Baumgart:2007,Veatch:2007}, it has been proposed
that inclusions (such as proteins) on such membranes are subject to 2D
analogs of critical FIF~\cite{Machta:2012fu}.  A notable advantage is
that 2D systems at criticality can be described by conformal field
theories (CFT)~\cite{Friedan:1984,Cardy:1989}: Casimir forces in a
strip are related to the central charge of the
CFT~~\cite{Cardy:1986fu,Kleban:1991ff,Kleban:1996pi}, with appropriate
modification for boundaries.  There are results for interactions
between circles~\cite{Machta:2012fu}, needles~\cite{Vasilyev::2012dz};
Ref.~\cite{Bimonte:2013cl} describes any compact shapes.  Here, we
consider the interaction between two wedges, or an array of wedges, as
depicted in Fig.~\ref{fig:geometry}. We show that (with appropriate
choice of CFT and boundary conditions) the FIF can be attractive or
repulsive depending on the angle of the wedge; and that arrangements
of stable equilibrium can be obtained with  truncated
wedges and arrays of them.

Consider two identically corrugated, infinite boundaries that enclose
a critical classical medium (e.g., a fluid or magnetic system at its
critical temperature $T_c$) described by a CFT.  The boundaries, $S_1$
and $S_2$, impose conformally invariant boundary conditions $a$
  and $b$, respectively, on the
medium. While our method is applicable to any shape, as specific
examples we study the periodic, wedge-like shapes in
Figs.~\ref{fig:geometry}(b,d).  As interactions at proximity are
dominated by the tips, we also consider the infinite wedges
depicted in Figs.~\ref{fig:geometry}(a,c).  Following our approach for
compact shapes~\cite{Bimonte:2013cl},
the strip with deformed boundaries is conformally mapped to a flat strip. 
Information about the intervening medium enters only
via its conformal charge $c$, and the scaling dimension
$h_{ab}$ of the boundary changing operator (BCO)
from $a$ to $b$; 
with $h_{ab}=0$ for like boundaries~\cite{Francesco:1997kx}. 
All information about the  shape of the deformed strip is encoded in the conformal map
to the flat strip. This map, and hence the FIF, can be obtained from the
solution to an electrostatic problem. 
In the following, we combine the normal ($y$) and lateral ($x$)
  components of the force into the complex expression $F=(F_x - i F_y)/2$.
For periodically deformed boundaries with wavelength $\lambda$ and length
$W\to\infty$, $F = F_\text{strip} + F_\text{geo} $, where the first
contribution is the force on a strip,
\footnote{Throughout the paper we measure energies in units of $k_B T_c$.}
\begin{equation}
  \label{eq:2}
  F_\text{strip} = - i \, \frac{\pi}{2} \left( \frac{c}{12} - \tilde
     \eta \right)   \frac{W}{\lambda} \frac{1}{2\ell^2} \int_{\Gamma_{cell}}
     (\partial_z w)^2 dz \,,
\end{equation}
that is determined by the free energy (per unit length) ${\cal
  F}_\text{strip}=-(\pi/2)(c/12-\tilde\eta) /\ell$ of a flat strip of width
$\ell$, and $\tilde\eta\equiv 2h_{ab}$. 
The second contribution is the geometric force
\begin{equation}
  \label{eq:3}
  F_\text{geo} = - \frac{i c}{24\pi} \, \frac{W}{\lambda}
  \int_{\Gamma_{cell}} \{ w,z \} dz \,,
\end{equation}
where  $\{w,z\} \equiv (\partial_z^3w/ \partial_z w) - (3/2)(\partial_z^2 w/ \partial_z w)^2$
is the Schwarzian derivative of the conformal map $w(z)$ of the deformed to the flat 
strip~\cite{Francesco:1997kx}. Due to  periodicity, it is sufficient to
construct $w(z)$ for a unit cell so that integrations in Eqs.~(\ref{eq:2},\ref{eq:3})
are restricted to a path
$\Gamma_\text{cell}$ that separates $S_1$ and $S_2$ within a unit cell
[cf. Fig.~\ref{fig:geometry}]. Of course, the forces are proportional
to the number of unit cells, $W/\lambda$. Whereas the strip force
depends on shape simply via the electrostatic capacitance \cite{Bimonte:2013cl}, the
geometric force has a more intricate dependence on the boundary
shapes. (The Schwarzian derivative vanishes if and only if $w(z)$ is a
global conformal map.) 

Conformal maps are physically realized as equipotential curves and
stream lines in electrostatics. We employ this analogy to derive
a general result for the Casimir force in terms of the electrostatic
potential $U(x,y)$ on the strip with the two boundaries held at
a fixed potential difference $\Delta U =1$.   
The conformal map is then given by $w(z)= U+iV$ where $V$ is
the conjugate harmonic function to $U$. Clealry $\ell=\Delta U=1$. 
Since Eqs.~(\ref{eq:2},\ref{eq:3}) 
involve only derivatives of $w(z)$, we use the Cauchy-Riemann 
equations to get $\partial_z w = \partial_x U - i \partial_y
U$ and eliminate $V$. For practical computations (e.g. using finite
element solvers) it is useful to express the Casimir force in terms of
line integrals of real valued vector fields that are fully determined
by derivatives of $U$. Parametrizing the contour $\Gamma_\text{cell}$ by
$\br(s)=[x(s),y(s)]$ for $0\le s\le 1$, and splitting into real and
imaginary parts, we obtain the force in terms of $c$, $\tilde\eta$ and $U$ as
\begin{widetext}
\vspace{-.5cm}
\begin{align}
  \label{eq:40a}
   F_\text{strip}  &= \frac{\pi}{2} \left( \frac{c}{12} - \tilde
     \eta \right)   \frac{W}{\lambda}  \left\{ 
\int_0^1 \bF_1[\br(s)] \cdot\br' (s) ds + i  \int_0^1 \bF_2[\br(s)]
\cdot\br' (s) ds \right\} \\
 \label{eq:40b}
 F_{geo} &= - \frac{i c}{24\pi} \, \frac{W}{\lambda} \left\{ 
\int_0^1 \bG_1[\br(s)] \cdot\br' (s) ds + i  \int_0^1 \bG_2[\br(s)] \cdot\br' (s) ds
\right\} \quad \text{with the vector fields}
\end{align}
\begin{align}
  \label{eq:45a}
\bF_1 & = 
\begin{pmatrix}
-\partial_x U \partial_y U\\
\frac{1}{2} \left((\partial_x U)^2-(\partial_y U)^2\right)
\end{pmatrix}, \quad
  \bF_2 = 
\begin{pmatrix}
-\frac{1}{2} \left((\partial_x U)^2-(\partial_y U)^2\right) \\
-\partial_x U \partial_y U
\end{pmatrix}\\
 \label{eq:45b}
  \bG_1 & = \frac{1}{\left((\partial_x U)^2+(\partial_y U)^2\right)^2}
\begin{pmatrix} 
\frac{1}{2} \left( \partial_x^2 U \partial_y U - \partial_x \partial_y
  U \partial_x U \right)^2 - \frac{1}{2} \left( \partial_x^2
  U \partial_x U +\partial_x \partial_y U \partial_y U\right)^2\\ 
\left( \partial_x^2 U \partial_x U +\partial_x \partial_y U \partial_y
U \right) \left( \partial_x^2 U \partial_y U - \partial_x \partial_y U
\partial_x U \right)
\end{pmatrix} \\
 \label{eq:45c}
  \bG_2 & = \frac{1}{\left((\partial_x U)^2+(\partial_y U)^2\right)^2}
\begin{pmatrix} 
- \left( \partial_x^2 U \partial_x U +\partial_x \partial_y U \partial_y
U \right) \left( \partial_x^2 U \partial_y U - \partial_x \partial_y U
\partial_x U \right)\\ 
\frac{1}{2} \left( \partial_x^2 U \partial_y U - \partial_x \partial_y
  U \partial_x U \right)^2 - \frac{1}{2} \left( \partial_x^2
  U \partial_x U +\partial_x \partial_y U \partial_y U\right)^2
\end{pmatrix}  \, .
\end{align}
\end{widetext}
We note that the strip force $F_\text{strip}$ is proportional to the
usual electrostatic force. This result also
implies  that the critical Casimir force between any pair
of mirror symmetric boundaries is attractive for $c>0$~\cite{Kenneth:2006wn,Bachas:2007vz}: In this case the
electrostatic potential must be constant along the $x$-axis of mirror
symmetry. Choosing this axis as $\Gamma_\text{cell}$ gives
$\br'(s)\sim -\hat{\mathbf x}$ and hence shows that both
$F_\text{strip}$ and $F_\text{geo}$ have a vanishing real part and a
negative imaginary part for $c/12-\tilde\eta >0$, which
includes like boundaries ($\tilde\eta=0$). This
implies a vanishing lateral force and positive normal force that
corresponds to attraction in our notation.

\begin{figure}[h]
\includegraphics[width= 1.\columnwidth]{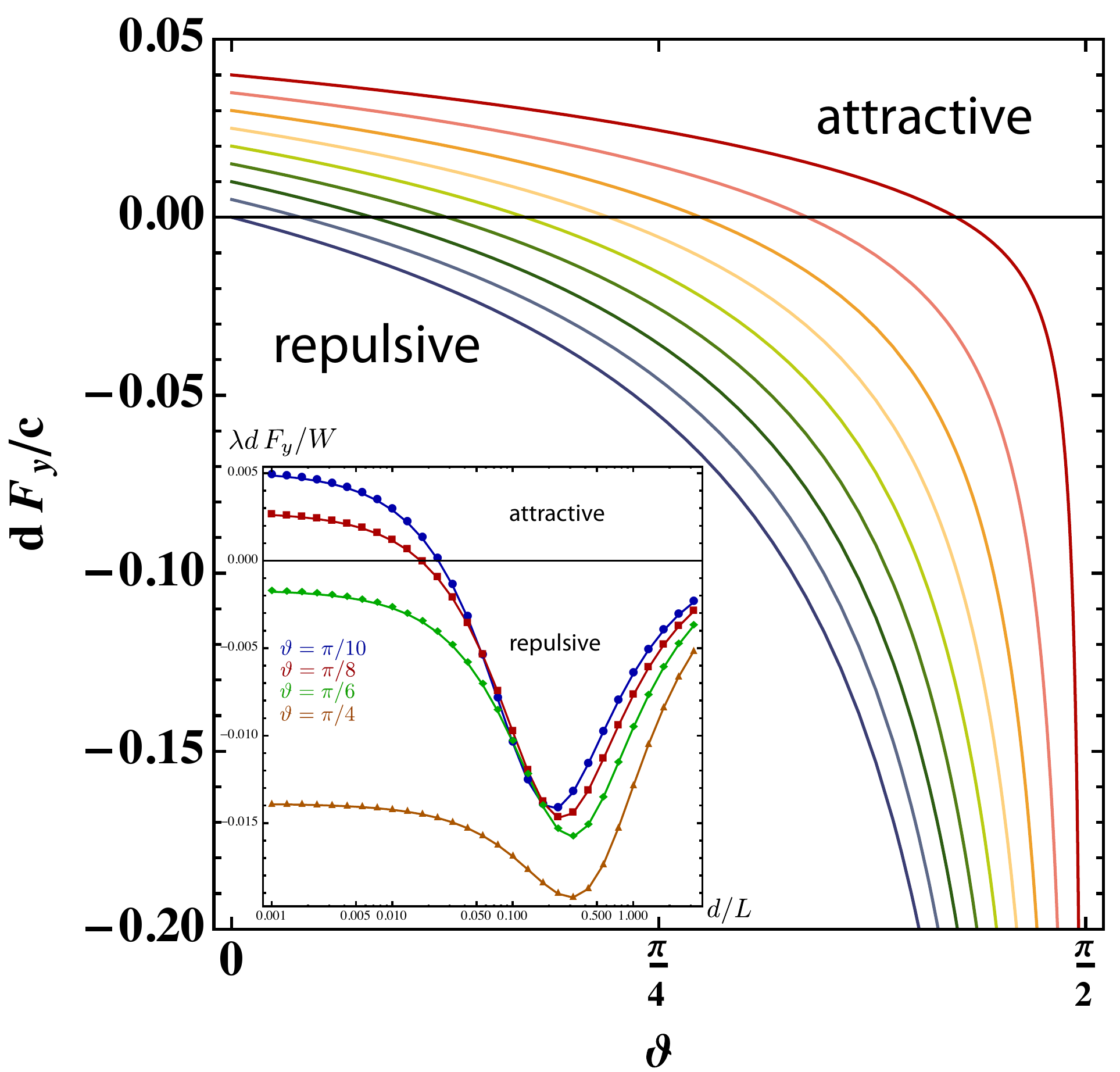}
\caption{Rescaled normal  force $F_y\sim 1/d$ between two
  infinite wedges as function of half opening angle
  $\vartheta$. The curves correspond to equidistant  CFT parameter ratios
  $\tilde\eta/c$ in the relevant interval  $1/12$ (red) to
  $1/8$ (blue). Inset: Rescaled  force  between triangular
  corrugations at selected opening angles for the tricritical Ising
  model and boundary changing operator $p=q=2$ with
  $\tilde\eta/c=3/28=0.1071$ (see text for details).}
\label{fig:force_wedges}
\end{figure}

Due to the simplicity of the related electrostatic problem, virtually
any boundary shape can be studied by computing $U$ either analytically
(e.g. using the Schwarz-Christoffel (SC) map for polygons
\cite{Smythe:1950vn}), or numerically (using finite element solvers). 
Here we consider a simple  profile  composed of a
periodic array of (truncated) wedges as in Figs.~\ref{fig:geometry}(b,d). 
For the triangular corrugations in Fig.~\ref{fig:geometry}(b), the SC map
yields an analytic result for the force in terms of a single parameter
implicitly determined in terms of $\vartheta$ and $L/d$. 
Due to lack of space, we delegate the full solution to a
forthcoming work, and study here short and large distances $d$ only.
At small $d \ll L$, the force is the sum of the contributions from
the tips of the wedges, such that the normal force $F_y = (W/\lambda)F_{\text{wedges},y}$.
The FIF between two infinite wedges  of opening angle 
$2\vartheta$ [Fig.~\ref{fig:geometry}(a)] is  proportional to $1/d$ on dimensional grounds, and given by
\begin{equation}
  \label{eq:37}
  F_{\text{wedges},y} = c \left[ \left(\frac{1}{12} - \frac{\tilde\eta}{c} \right) 
    \frac{\pi}{\pi-2\vartheta} +\frac{1}{24} \frac{\pi-2\vartheta}{\pi-\vartheta}
\right] \frac{1}{d} \, ,
\end{equation}
where the first term corresponds to $F_\text{strip}$ and the second to
$F_\text{geo}$. The amplitude of this force is shown in
Fig.~\ref{fig:force_wedges} for different values of 
$\tilde\eta/c$, corresponding to unlike boundary conditions ($\tilde\eta>0$). 
Interestingly, for $\tilde\eta/c < 1/8$ the
force becomes attractive below a critical opening angle $\vartheta$
\footnote{The threshold $\tilde\eta/c = 1/8$ corresponds to a free
  field theory $(c=1)$ with mixed Dirichlet and Neumann boundary
  conditions.}.  This is different from the asymptotic large distance
force between the boundaries, $F_y = (\pi/2)(c/12 -
\tilde\eta)W/d^2$, which is repulsive for $\tilde\eta/c >
1/12$. Hence, there is a reversal of the force between triangular
corrugations from repulsive to attractive with decreasing separation
$d$ if 
\begin{equation}
\label{eq:stab_cond}
\frac{1}{12} < \frac{\tilde\eta_{(ab)}}{c} = \frac{2h_{(ab)}}{c} < \frac{1}{8}\,,
\end{equation}
and $\vartheta$ is sufficiently small.  This is confirmed by our full
analytic solution at all distances which is shown in the inset of
Fig.~\ref{fig:force_wedges} for different opening angles and a BCO of
the tricritical Ising model that obeys Eq.~\eqref{eq:stab_cond}.
The change of sign corresponds to an {\it unstable} point.

However, these results together with the expected validity of the
proximity force approximation (PFA) at very short separations suggest
the possibility of a stable point if the tips of the wedges are
truncated and replaced by plateaus of width $H \ll  L$, as in
Fig.~\ref{fig:geometry}(d). Indeed, for a single pair of truncated
wegdes [Fig.~\ref{fig:geometry}(c)], PFA at short distances $d\ll H$ suggests
\begin{equation}
  \label{eq:trunc_wedges_pfa}
  F^\text{PFA}_{\text{tr.~wedges},y} = \frac{\pi}{2} \left( \frac{c}{12} -
    \tilde\eta \right) \frac{H}{d^2} \, ,
\end{equation}
which is repulsive for $\tilde\eta/c>12$.  At distances $ d\gg H$, the
plateaus become irrelevant, and the force approaches the result of
Eq.~\eqref{eq:37}. Hence two truncated wedges must have a stable point
at intermediate distance if Eq.~\eqref{eq:stab_cond} holds and
the opening angle is sufficiently small. This expectation is confirmed
by an exact computation (using a SC map) of the normal and lateral
force between two truncated wedges with lateral shift $\delta$ [see
Fig.~\ref{fig:geometry}(c)]. We additionally confirm that
this configuration is stable with respect to displacements in the lateral direction.

\begin{figure}[h]
\includegraphics[width= .95\columnwidth]{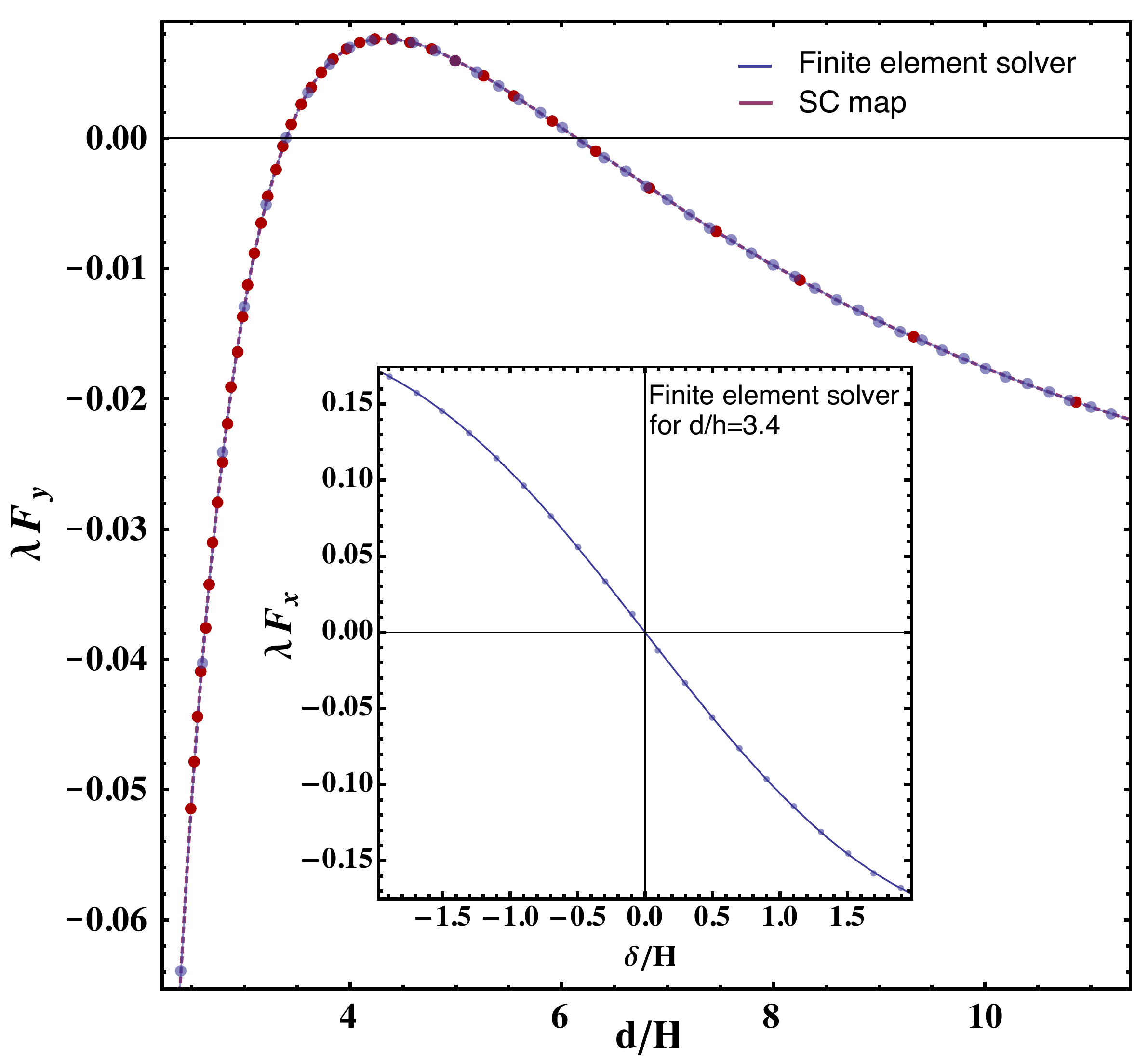}
\caption{The normal Casimir force acting on boundary ($S_2$) for
  vanishing lateral shift $\delta$, as function of separation $d$
  for the geometry of Fig.~\ref{fig:geometry}(d) with
  $\vartheta=\pi/10$. Parameters applicable to the tricritical Ising model with
   boundary changing operator of scaling dimension
  $h_{2,2}=3/80$ are used. Inset: The lateral Casimir force close to stable
  separation $d=3.4 \, H$ as a function of the scaled lateral shift.}
\label{fig:fig:norma_lateral_force_TIM}
\end{figure}

\begin{figure}[tbh]
\includegraphics[width= .85\columnwidth]{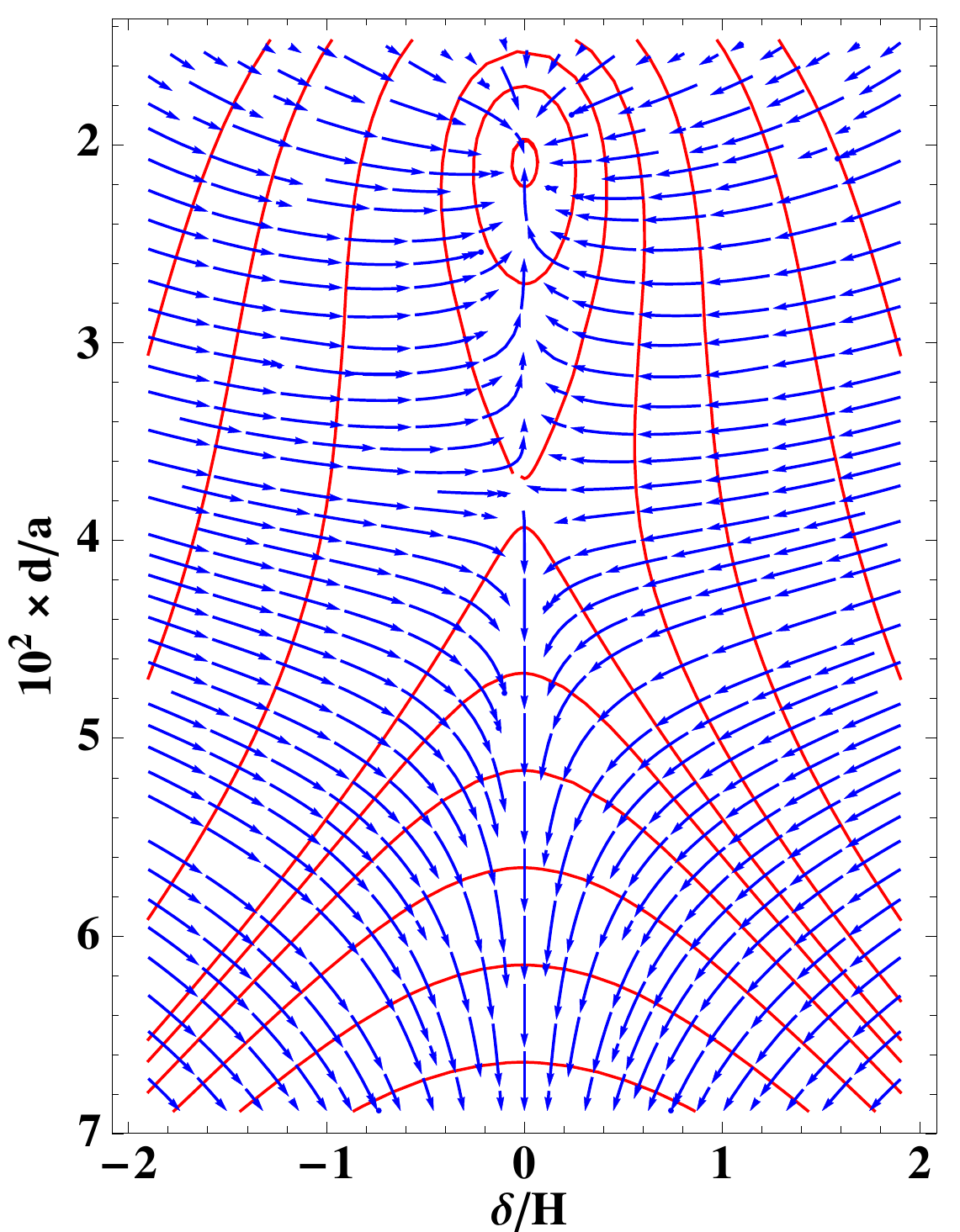}
\caption{Casimir force field and curves of constant Casimir potential 
as function of normal separation $d$ and lateral shift $\delta$,
for the same geometry and CFT as in  Fig.~\ref{fig:fig:norma_lateral_force_TIM}. 
The saddle point, and the stable equilibrium point are clearly visible.}
\label{fig:force_field_TIM}
\end{figure}

Combining the above findings, for the truncated triangular
corrugations of Fig.~\ref{fig:geometry}(d) we expect under the
condition \eqref{eq:stab_cond} and for sufficiently small $\vartheta$
a stable equilibrium point (in both directions), and  a saddle point at larger separations. 
For this geometry, a SC transformation can be performed in
principle, but has to be evaluated numerically (which is cumbersome for
a finite lateral shift $\delta$). Hence, we employ the analogy to
electrostatics as described by Eqs.~(\ref{eq:40a},\ref{eq:40b}).
The electrostatic potential is computed by a finite element solver
(FES) and subsequently the resulting vector fields [see
Eqs.~(\ref{eq:45a}-\ref{eq:45c})] are integrated along the contour
$\Gamma_\text{cell}$. To be specific, we chose the unitary CFT with
conformal charge $c=7/10$ that describes the tricritical Ising model
(TIM) and chose boundary conditions that are connected by the BCO with
scaling dimension $h_{ab}=3/80$ so that condition in Eq.~\eqref{eq:stab_cond}
is fulfilled. (The reason for these particular choices shall become
clear below when we discuss possible CFT's.) 
The accuracy of the FES can be established by
comparing its results to those  from a SC transformation for
vanishing lateral shift $\delta$. The results of both methods for the
normal force (acting on the lower boundary $S_2$) between truncated
triangular corrugations with $\vartheta=\pi/10$ are shown in
Fig.~\ref{fig:fig:norma_lateral_force_TIM}. The agreement is excellent
and confirms sufficient accuracy of the FES. The normal force shows
the expected sign reversals from repulsive to attractive and back to
repulsive.  For a finite lateral shift $\delta$ we used the FES to
compute both normal and lateral force. The lateral force at a normal
distance close to the stable point ($d/H=3.4$) is plotted in the inset
of Fig.~\ref{fig:fig:norma_lateral_force_TIM}, demonstraing mechanical
stability also in the lateral direction. The global force field and curves of
constant Casimir potential are shown in Fig.~\ref{fig:force_field_TIM}. 
The presence of  a stable equilibrium point, and a
saddle point, (both at $\delta=0$) is clearly confirmed. 

It is interesting to explore which critical systems allow for a stable equilibrium
point in the above geometry.
In the following, we identify unitary minimal CFT models with $c<1$ 
that permit boundary conditions
consistent with the criteria in Eq.~\eqref{eq:stab_cond}.  
For unitary models the conformal charge is restricted to the discrete
values $c=1-6/[m(m+1)]$ with integer $m\ge 2$. The allowed scaling
dimensions of the primary operators can assume the values \cite{Friedan:1984}
\begin{equation}
  \label{eq:hs}
  h_{p,q} = \frac{[(m+1)p-mq]^2-1}{4m(m+1)} \, ,
\end{equation}
with $p=1,2,\cdots, m-1$, and $q=1,2, \cdots p$. 
Cardy has shown that all
possible highest weight states with scaling dimension $h_{p,q}$ may be
realized by a BCO for an appropriate choice of boundary conditions
$(ab)$ on the flat strip~\cite{Cardy:1984qf}. 
The two conformally invariant boundary conditions, or states that are connected 
by a BCO, are determined by the fusion rules for the two (bulk) primary fields
that correspond to the boundary states. 
The condition of Eq.~\eqref{eq:stab_cond} can only be fulfilled for $p=q$. 
It turns out that for $m=3$ (Ising model), $m=5$ (3-state Potts model), and $m=7$ 
no primary operator obeys the condition. 
For all other models with $m \le 13$ there is exactly one operator whose 
dimension obeys the condition, while for $m>13$ two
or more operators with suitable dimensions may exist. 
The simplest models with  suitable BCO's are the TIM ($m=4$,
$c=7/10$, $h_{2,2}=3/80$) and the tricritical 3-state Potts model
($m=6$, $c=6/7$, $h_{3,3}=1/21$). 
These considerations underlie our choice of the TIM for Figs.~\ref{fig:fig:norma_lateral_force_TIM},~\ref{fig:force_field_TIM}.
There are certainly other minimal models that also allow for conformally 
invariant boundary conditions that lead to a stable point. 

Tricritical points are a common feature of many phase diagrams, corresponding
to the point where a continuous transition becomes first order,
as can be observed by addition of vacancies or other impurities to an Ising magnet,
or helium 3 to superfluid helium 4 (see, e.g. Ref.~\cite{Lawrie:1984xp} for a review).
Possible boundary conditions compatible with a renormalization group fixed
point at tricriticality are discussed in Ref.~\cite{Krech:1992}. 
Fusion rules in CFT~\cite{Francesco:1997kx} provide another route to 
characterizing conformally invariant boundary conditions.
For the TIM with the BCO of scaling dimension $h_{2,2}=3/80$,
the relevant fusion rules are $|1/10\rangle \times |7/16 \rangle =
|3/80\rangle$, $|3/5\rangle \times |7/16 \rangle = |3/80\rangle$,
$|3/2\rangle \times |3/80 \rangle = |3/80\rangle$, $|0 \rangle \times
|3/80 \rangle = |3/80\rangle$, $|1/10 \rangle \times |3/80\rangle = |3/80 \rangle + |7/16\rangle$, and
$| 3/5\rangle \times |3/80\rangle = | 3/80\rangle + |7/16\rangle$.
The last two rules are relevant since for a semi-infinite strip the
lowest dimension determines the free energy. 
Through appropriate choice of surface couplings and magnetic fields,
the TIM admits the following conformally invariant boundary
states~\cite{Chim:1996fr,Affleck:2000dz}: (i) A disordered state of
free spins, corresponding to $|7/16\rangle$; (ii) Maximally ordered
(fixed) spins ($+$ or $-$), with $|0\rangle$, $|3/2\rangle$.  The
phase transition between the above surface states can occur through
(iii) Partially polarized ($+$ or $-$ with vacancies) at finite
surface fields, for $|1/10\rangle$, $|3/5\rangle$; or (iv) Through 
a so called degenerate point at zero surface field, with
$|3/80\rangle$.  The fusion rules show that a stable point with
vanishing FIF can occur for the following combinations $(a,b)$ of
boundary conditions:
\begin{enumerate}
\setlength{\itemsep}{-3pt}
\item (fixed spin, degenerate),
\item (partially polarized, free spin),
\item (partially polarized, degenerate).
\end{enumerate}
The stability of these boundary states (fixed points) with respect to
a boundary magnetic field and spin couplings is determined by the
boundary phase diagram of the TIM~\cite{Affleck:2000dz}.  The free and
fixed boundary conditions can be achieved relatively easily (at least
in simulations), while the degenerate and partially polarized states
require tuning one parameter (the surface coupling, or surface field).
We expect that the combination of {\it partially polarized} and {\it
  free spin} conditions is the most promising candidate.

The conditions obtained here for the observation of a stable equilibrium
point with FIF are rather restrictive. 
This demonstrates on the one hand the difficulty of achieving stability
solely by FIF, on the one hand, and absence of its strict impossibility 
(ala Earnshaw~\cite{Rahi:2010yg}) on the other.
For other examples in 2D, we could look for other realizations of CFT
in interface (restricted solid-on-solid) models~\cite{Pasquier:1987}.
It would be quite interesting to explore the possibility of stability with
critical FIF in three dimensions. 
It is not a priori clear if the necessary conditions in higher dimensions
will be less or more restrictive. 
Since $d=3$ is the upper critical dimension for the TIM, at least in this case
the question could in principle be resolved by generalizing standard
field theory methods~\cite{Krech:1992} to wedge/cone 
geometries~\cite{Maghrebi:2011bk,Maghrebi:2011}.

We thank R.~L.~Jaffe for valuable discussions.  This research was
supported by Labex PALM AO 2013 grant CASIMIR, and the NSF through
grant No. DMR-12-06323.


\end{document}